\documentclass[pra, superscriptaddress, twocolumn]{revtex4}

\usepackage{subfigure}
\usepackage{empheq}

\usepackage{dcolumn}
\usepackage{amsmath,amssymb}
\usepackage{bm}
\usepackage{color}
\usepackage{overpic}
\usepackage{latexsym}
\usepackage{epstopdf}
\usepackage{color}
\usepackage[english]{babel}
\usepackage{graphicx}
\usepackage{psfrag} 
\usepackage{epsf} 
\usepackage{subfigure} 
\usepackage{amsfonts}
\usepackage{bm}
\usepackage{epstopdf}\usepackage{appendix}

\definecolor{mygrey}{gray}{0.35}
\definecolor{myblue}{rgb}{0.2,0.2,0.8}
\definecolor{myzard}{cmyk}{0,0,0.05,0}
\definecolor{mywhite}{rgb}{1,1,1}
\definecolor{myred}{rgb}{1,0.,0.3}

\usepackage[colorlinks=true,citecolor=myblue,linkcolor=myred]{hyperref}
\def\beq{\begin{equation}}
\def\eeq{\end{equation}}
\def\ba{\begin{align}}
\def\enda{\end{align}}
\def\bi{\begin{itemize}}
\def\ei{\end{itemize}}

 \newcommand{\ket}[1]{|#1\rangle}
 \newcommand{\bra}[1]{\langle #1|}
 \newcommand{\braket}[2]{\langle #1|#2\rangle}

\usepackage[normalem]{ulem}

\begin{document}

\title{On the description of composite bosons in discrete models}
\author{Paula C\'espedes}
\affiliation{FAMAF, Universidad Nacional de C\'{o}rdoba,
Ciudad Universitaria, X5016LAE, C\'{o}rdoba, Argentina}
\author{Elena Rufeil-Fiori}
\affiliation{Instituto de F\'{i}sica Enrique Gaviola, CONICET and Universidad 
Nacional de C\'{o}rdoba,
Ciudad Universitaria, X5016LAE, C\'{o}rdoba, Argentina}
\author{P. Alexander Bouvrie}
\affiliation{Centro Brasileiro de Pesquisas F\'isicas, Rua Dr. Xavier Sigaud 
150, Rio de Janeiro, RJ 22290-180, Brazil}
\author{Ana P. Majtey}
\affiliation{Instituto de F\'{i}sica Enrique Gaviola, CONICET and Universidad 
Nacional de C\'{o}rdoba,
Ciudad Universitaria, X5016LAE, C\'{o}rdoba, Argentina}
\author{Cecilia Cormick}
\affiliation{Instituto de F\'{i}sica Enrique Gaviola, CONICET and Universidad 
Nacional de C\'{o}rdoba,
Ciudad Universitaria, X5016LAE, C\'{o}rdoba, Argentina}

\date{\today}

\begin{abstract}
The understanding of the behaviour of systems of identical composite bosons has 
progressed significantly in connection with the analysis of the entanglement 
between constituents and the development of coboson theory. The basis of these 
treatments is a coboson ansatz for the ground state of a system of $N$ pairs, 
stating that in appropriate limits this state is well approximated by the 
account of Pauli exclusion in what would otherwise be the product state of 
$N$ independent pairs, each described by the single-pair ground state. In this 
work we study the validity of this ansatz for particularly simple problems, and 
show that short-ranged attractive interactions in very dilute limits and a 
single-pair ground state with very large entanglement are not enough to render 
the ansatz valid. On the contrary, we find that the dimensionality of the 
problem plays a crucial role in the behaviour of the many-body ground state.
\end{abstract}

\maketitle

\section{Introduction}

The fact that composite systems made up of an even number of fermionic 
constituents behave in practice like elementary bosons is long known \cite{fierz_39,pauli_40,Annett_book}, but a 
rigorous understanding of this behaviour has only been gained recently. Such 
advancement has taken place including elements of quantum information theory, 
in particular entanglement theory \cite{law_2005,chudzicki_2010}, but has been specially accomplished by the 
development of coboson theory \cite{combescot_2008,combescot_2011,combescot_2015}. 
This formalism, originally designed to study phenomena such as excitations in a 
crystalline solid by means of excitons or superconductivity through Cooper 
pairs, dealing with pairs of fermions as compound bosons \cite{combescot_2015}, 
was also successfully applied to a larger variety of systems 
\cite{shiau_2016}, including molecular Bose-Einstein condensates 
in ultra-cold interacting Fermi gases \cite{combescot_2008b, combescot_2016, 
bouvrie_2017, bouvrie_2018}. Manifestations of the effects of Pauli 
exclusion in composite bosons 
made of fermions have been analyzed for thought interference experiments in  
\cite{tichy_2012, bouvrie_2016} and for potential implementations with 
condensates in \cite{combescot_2017}. We note, however, that although the coboson treatment 
provides a good approach for the description of Feshbach molecules 
\cite{combescot_2008b, combescot_2016, bouvrie_2017}, its application for the understanding of 
atomic Cooper pairs is non-trivial \cite{pong_law_2007}, and it remains unclear 
whether the description of the BEC-BCS crossover in terms of coboson theory is 
possible. 

Precisely because of the relevance of coboson theory and its success in 
describing 
several physical phenomena, an understanding of its regime of validity is 
specially desirable. In particular, it is not evident when a key element of the 
theory, namely the so-called coboson ansatz for the ground state, provides an 
appropriate description of the zero-temperature state of a system of $N$ pairs. 
The ansatz approximates the ground state of $N$ composite bosons by a state 
which is given by the repeated action over the vacuum of an operator creating 
one pair in the single-pair ground state, including a proper additional 
normalization to account for the effect of Pauli exclusion.
Loosely speaking, one expects this ansatz to be valid when the constituents 
interactions are sufficiently short ranged, the system is sufficiently dilute, 
and the ground state for a single pair is highly entangled 
\cite{combescot_2015,combescot_2016}.

In this work we show in a particularly simple example that these conditions are 
not sufficient for the ansatz to be valid, and that the dimensionality of the 
problem actually plays a key role. This is due to the fact that one-dimensional 
models, even with short-ranged interactions, can lead to long-range 
correlations in the ground states of several pairs. It is important to stress 
that coboson theory was not developed for one-dimensional problems, and that 
our observations do not undermine the importance of the theory. On the 
contrary, we expect to contribute to the usefulness of this theoretical body of 
work by helping to establish more clearly its limits of applicability.

In particular, the model we consider was motivated by the one introduced in \cite{tichy_2012}, where the coboson ansatz was taken as initial state to study the 
effects of compositeness in a thought interference experiment. Although the use 
of the coboson ansatz in \cite{tichy_2012} is not at all essential for the 
analysis 
presented, the article can convey the mistaken impression that the ansatz is 
valid for the system considered, namely a 1D chain of discrete sites along 
which the constituent fermions can hop, and including a short-ranged 
interaction between fermions of different species such that pairs are always 
strongly bound. 

Here, we analyze this model in the situation where the coboson ansatz is 
expected to work best, namely for maximum entanglement between constituents, 
and focusing on the simplest case of two pairs. We show that as the particle 
density becomes lower, the coboson state does not approach the true ground state 
of the system, and indeed the fidelity decreases reaching a limiting value of 
$8/\pi^2$. We also show that extending this model to a ``ladder'' with a fixed width 
of $n$ sites, to allow pairs to cross each other, does not significantly modify 
this result. The reason for this behaviour is 
the long-range character of the correlations appearing for one-dimensional 
settings, which cannot be captured by the ansatz. On the contrary, when 
the system is made truly two-dimensional, the coboson ansatz becomes a good 
description of the system as long as it is dilute enough.

This article is organized as follows: In Sec. \ref{sec:cobosons} we introduce 
briefly some very basic elements of the coboson formalism. In Sec. 
\ref{sec:model} we 
give the details of the model under consideration, the single-pair ground 
manifold and the effective Hamiltonian for the ground manifold of $N$ pairs. 
In Sec. \ref{sec:1D comparison} we give the exact solution for the ground state 
of two pairs and compare it with the coboson ansatz, observing that the 
fidelity with the true ground 
state decreases as the number of sites is increased. Section \ref{sec:2D} 
presents 
an analysis of the two-dimensional generalization of the system, showing that 
the coboson ansatz behaves well in this case, and in Sec. \ref{sec:conclusion} 
we provide a discussion of the results and summarize our conclusions.

\section{The coboson ground state}\label{sec:cobosons}

We consider a system of identical composite bosons, each made of two distinguishable 
fermions. This section provides a brief overview of the coboson ansatz for the 
ground state of $N$ such pairs. For a more complete 
introduction to coboson theory, we refer the reader to \cite{combescot_2008,combescot_2003,law_2005,chudzicki_2010,combescot_2015}.
For a given Hamiltonian corresponding to a single pair, the ground state $|\psi\rangle$
defines the coboson creation operator $c^\dagger$, namely the operator which acts on 
the vacuum creating a single pair in the ground state, $|\psi\rangle=c^\dagger|0\rangle$. 
In the Schmidt 
basis, this operator can be written as \cite{law_2005}
\beq 
 c^\dagger=\sum_{\alpha=1}^S \sqrt{\lambda_\alpha} \,  a^\dagger_\alpha 
 b^\dagger_\alpha 
\eeq
with
\beq
\lambda_1 \ge \lambda_2 \ge \dots  \ge 0
\eeq
the Schmidt coefficients satisfying
\beq
\sum_{\alpha=1}^S \lambda_\alpha =1
\eeq
and $S$ the (finite or infinite) Schmidt rank. The operators 
$a^\dagger_\alpha$, 
$b^\dagger_\alpha$ create one particle of kind $a$ or $b$ respectively in 
the corresponding Schmidt modes $\alpha$. The operator 
$c$ obeys the following commutation 
relations 
\begin{eqnarray}
[ c , c] &=& [ c^\dagger , c^\dagger] = 0 , \nonumber \\ {} \label{ccdagger}
[ c , c^\dagger] &=& 1-\Delta ,
\end{eqnarray}
where 
\begin{equation}
\Delta = \sum_{\alpha=1}^S \lambda_\alpha ( a_{\alpha}^\dagger  a_{\alpha} +  
b_{\alpha}^\dagger  b_{\alpha})\,.
\end{equation}

One can write a normalized state of $N$ 
composite bosons obtained after acting $N$ times with the coboson creation 
operator in the form \cite{law_2005,combescot_2008}
\begin{equation}
\ket{N} = \frac{\left( c^\dagger \right)^N}{\sqrt{ N! \chi_{N} } } \ket{0}.
\label{eq:ansatz}
\end{equation}
where $\chi_{N}$ is the compositeness 
normalization factor \cite{law_2005, 
combescot_2003, combescot_2008, tichy_2014} which depends on the Schmidt 
coefficients and which accounts for the sub-normalization of the state resulting of adding 
$N$ bi-fermions 
to the vacuum. For composite bosons made of two fermions the normalization 
factor takes the form \cite{law_2005},
\begin{equation}\label{chiN}
\chi_{N} = N! \sum^S_{p_N > p_{N-1} > \text{...}> p_1} \lambda_{p_1} 
\lambda_{p_2} \text{...}\lambda_{p_N}
\end{equation}
which is the elementary symmetric polynomial 
\cite{macdonald_book}.
The idea that the state $\ket{N}$ provides a good approximate description of the ground 
state of a system of $N$ cobosons is a key element of coboson theory, and we 
refer to this in the following as the coboson ansatz. 

For the case $N=2$, the normalization coefficient is equal to $\chi_2 = 1-P$, 
with $P=\sum_{\alpha}\lambda_\alpha^2\leq 1$ the purity of the reduced density matrix of one of the 
constituent 
particles of a pair in the ground state $|\psi\rangle$. In general, the behaviour of pairs as 
approximate elementary bosons can be related with the normalization 
coefficients, and bosonic behaviour is recovered when 
$\chi_N/\chi_{N-1}\simeq 1$ \cite{law_2005,combescot_2011b,tichy_2012b,tichy_2014}.

\section{The model} \label{sec:model}

The problem we consider is a one-dimensional array of $L$ sites with two species 
of fermions that can hop along them. Fermions 
of different species have a very strong attraction, so that if the numbers of particles of
both species are equal then the low energy manifold has all particles in pairs. 
More precisely, the Hamiltonian takes the form:
\begin{multline}
H = -U_0 \sum_{j=1}^L a_j^\dagger a_j \, b_j^\dagger b_j \\+ \frac{J}{2} 
\sum_{j=1}^L (a_j^\dagger a_{j+1} +
b_j^\dagger b_{j+1} + {\rm H.c.} )
\label{eq:H}
\end{multline}
where $a_j$ ($a_j^\dagger$) destroys (creates) a particle of type $a$ in site 
$j$, $b_j$ ($b_j^\dagger$) does the same for a particle of type $b$, and we 
assume for definiteness that the operators associated with particles of 
different kind commute (the results are the same if they
anticommute \cite{combescot_2015}).

We consider that the interaction energy is much stronger than the hopping, i.e. 
$U_0\gg J$. We then study this problem analytically using perturbation theory. 
As will be shown in the following, the restriction to the limit 
when particles always tunnel in pairs makes our system an instance of the 
hard-core Bose-Hubbard model, which is equivalent to the Heisenberg model 
\cite{baxter_book, bethe_1931, karabach_1997, Blundell_book}. We note that this limit of 
very strongly bound pairs is the one studied in \cite{tichy_2012}, and it is also particularly 
relevant for our interests since it is the situation where the coboson description should be most 
appropriate. We remark that the problem considered in \cite{tichy_2012} includes also 
site-dependent energies, as a free parameter to control the amount 
of entanglement in the single-pair ground state. Here for simplicity we focus 
on the case where the coboson ansatz is supposed to work best, namely when 
the entanglement is maximum. This corresponds to the translation-invariant case with all site 
energies equal and periodic boundary conditions. In the following we will derive 
the effective Hamiltonian for the lowest-energy manifold in this model, for a single pair and for 
$N$ pairs.

\subsection{Single-pair basis - Ground manifold}

The Hilbert space of a single pair of particles, one of each kind, divides into 
a ground manifold composed by the states where the particles are paired (i.e. 
occupying the same site), containing $L$ states, and an excited manifold where 
the particles are not paired, with dimension $L^2-L$. The ground manifold 
energy, to zero order in the hopping, is $-U_0$, while the excited manifold has 
zero energy up to same order. Using perturbation theory, we can find the 
approximate eigenstates within each of these highly degenerate manifolds.

As already explained in \cite{tichy_2012}, to first order the hopping Hamiltonian for 
the ground manifold vanishes, so that the first non-zero correction is of second 
order and has the form:
\beq
H_g \simeq -U_0 - P_g \frac{H_J^2}{U_0} P_g
\eeq
where $P_g$ is the projector onto the ground manifold, and $H_J$ is the hopping 
part of the Hamiltonian. It is straightforward to see that this gives (for the 
case of a single pair):
\beq\label{eq:Hg1}
H_g^{(1)} \simeq -U_0-\frac{J^2}{U_0} - \frac{J^2}{2U_0}\sum_{j=1}^L 
(\ket{j,j}\bra{j+1,j+1} +H.c.)
\eeq
so that particles always tunnel in pairs. We note that the form of this effective Hamiltonian is 
independent of the sign of $J$; indeed, a change of sign in $J$ in the original Hamiltonian 
\eqref{eq:H} can be reabsorbed by a sign flip in the creation and annihilation operators 
corresponding to all odd (or all even) sites, and this sign flip becomes irrelevant when only pairs 
can tunnel.

It is also easy to diagonalize this Hamiltonian with a Fourier transformation. The coboson operators which create the single-pair eigenstates within this manifold are thus found to be of the form:
\beq
c^\dagger_k = \frac{1}{\sqrt{L}} \sum_j e^{-2\pi i kj/L} a^\dagger_j b^\dagger_j
\label{eq:Fourier}
\eeq
with corresponding energies:
\beq
E_k = -U_0 -2\frac{J^2}{U_0} \cos^2(k\pi/L)
\eeq 
where $k$ runs from 0 to $L-1$. Except for the lowest state within this 
manifold (and for even $L$ also the highest), 
the eigenstates are doubly degenerate.

The single-pair ground-state energy is thus:
\beq
E_0 = -U_0 -2\frac{J^2}{U_0} \,,
\label{eq:energy one pair}
\eeq 
and the ground state is:
\beq\label{eq:phi_0}
\ket{G}_{N=1} = \frac{1}{\sqrt{L}} \sum_j \ket{j,j} \,,
\eeq
with the ground-state coboson creation operator given by
$c^\dagger = c_0^\dagger$. This greatly simplifies some calculations, because 
all Schmidt coefficients of this state are equal to $1/L$, and it is 
straightforward to compute the form of the coboson ansatz for 
the ground state of $N$ pairs and the corresponding normalization factors (see 
Sec. \ref{sec:1D comparison}). We note also that in this case $S=L$, and the entanglement between 
the components of a single pair in the ground state is characterized by the purity $P=1/L$, 
corresponding to a maximally entangled state for each fixed number of sites.

\subsection{Effective Hamiltonian for the ground manifold of $N$ pairs}

Following similar lines as before, one can use perturbation theory for the 
effective Hamiltonian of the ground manifold for the case of several pairs. Once 
more, it is trivial to see that the ground manifold is formed by the states in 
which all particles are paired, and to zero order in the hopping the energy of 
this manifold is $-NU_0$. In the following we analyze the corrections when 
second-order terms in the hopping are introduced.

In analogous manner as in the single-pair case, the effective Hamiltonian for 
the ground manifold takes the form:
\beq
H_g \simeq -N U_0 - P_g \frac{H_J^2}{U_0} P_g
\eeq
Now the projection to the ground manifold of the term proportional to $H_J^2$ has 
an additional term coming from the fact that hopping of a particle out of a 
given site might be forbidden if there is already a particle sitting there. 
This leads to the form:
\begin{multline}
H_g^{(N)} \simeq -N \left(U_0+ \frac{J^2}{U_0}\right) + \frac{J^2}{U_0} \sum_j 
N_j N_{j+1} \\ - \frac{J^2}{2 U_0} \sum_j (T_j^+ + T_j^-)
\label{eq: eff H}
\end{multline}
where $N_j$ is the number of pairs in site $j$, and $T_j^{\pm}$ are the 
operators that correspond to hopping of a pair from site $j$ to $j\pm1$.
For the case $N=1$ this clearly reduces to the single-pair effective 
Hamiltonian of the ground manifold given by (\ref{eq:Hg1}). 

One can see from the form of the ground-manifold Hamiltonian for 
$N$ pairs that, as in the single-pair case, there is a hopping term that will 
tend to delocalize the cobosons; but now there is an additional interaction 
between sites that will compete with the hopping. This interaction is 
repulsive, and therefore one expects that the ground state will have 
delocalized pairs but which are unlikely to be found next to each other. The 
exact ground state of the effective Hamiltonian and its energy for the case of two pairs are discussed in Sec. \ref{sec:1D comparison}.

\subsection{Relation with the Heisenberg model}

Discrete hard-core boson models are equivalent to 
spin-$1/2$ systems, and indeed Hamiltonian (\ref{eq: eff H}) is equivalent to the 
Heisenberg Hamiltonian 
for a chain of spins $1/2$, by means of the identifications: $a_j b_j \equiv \sigma_j^-$, the spin lowering operator,  $N_j 
\equiv (\sigma_j^z+1)/2$. 
Tunneling terms of the form $T_j^+ + T_j^-$ 
can then be written as interactions of the form $\sigma_j^x \sigma_{j+1}^{x} + 
\sigma_j^y \sigma_{j+1}^{y}$, while the term of the form $N_j N_{j+1}$ 
corresponds to an interaction through $\sigma_z$ plus a global field along $z$ 
direction.In order to obtain the Heisenberg 
Hamiltonian one has to additionally apply on every other spin a rotation about the $z$ axis 
in order to flip the sign of the corresponding $\sigma_x$ and $\sigma_y$ operators. This only works 
for an even number of sites, but for big systems we do not expect the parity of the number 
of sites to play a crucial role. 

The
effective ground-manifold Hamiltonian then takes the form:
\beq
H_g^{(N)} \equiv -N U_0 + \frac{J^2}{4U_0} 
\left(H_H - L \right)
\eeq
where $H_H$ is the Heisenberg Hamiltonian:
\beq
H_H = \sum_j \vec{\sigma}_j \cdot \vec{\sigma}_{j+1} \,.
\eeq
The ground-state energy for the system of $N$ pairs 
can be obtained from the minimum 
energy of the Heisenberg Hamiltonian in the manifold corresponding to $N$ 
particles, which fixes the total projection of the spin
along $z$, $\sigma_T^z = 2N-L$. 

This means the problem can be approached with the Bethe 
ansatz \cite{karabach_1997}, and its properties have been studied extensively \cite{Blundell_book}. In general, it is not possible to find an exact analytical 
solution of the Heisenberg model for arbitrary values of $N$. However, for 
the particular case with two particles only, an exact solution can be written 
for the ground state, as will be discussed in Sec. \ref{sec:1D comparison}.

\section{Exact ground state for two pairs, and comparison with the coboson 
ansatz} \label{sec:1D comparison}

For the case of two pairs, the exact analytical solution of the effective 
Hamiltonian is known and has the form \cite{karabach_1997}:
\beq
\ket{G}_{N=2} = A \sum_{j_1 < j_2} \sin\left[\pi \frac{d(j_1, j_2)-1/2}{L-1} 
\right] \ket{j_1, j_2}
\label{eq:true ground state}
\eeq
where $A$ is a normalization factor and $d(j_1, j_2)$ is the distance between 
the two occupied sites, taken mod $L$. 
It is convenient to notice that although the 
mapping to the Heisenberg Hamiltonian as described above was valid for even 
$L$ only, this expression for the ground state holds also for $L$ odd.

The coboson ansatz for the ground state of a system of many particles, when
the numbers of particles of each kind are both equal to $N$, is given by 
Eq.~(\ref{eq:ansatz}). Given the form of the operator $c_0^\dagger$ in 
(\ref{eq:Fourier}), the ansatz for this problem leads to:
\beq
\ket{N} = 
\left[
\begin{pmatrix}
L \\ N
\end{pmatrix}
\right]^{-1/2}
\sum_{j_1<j_2<\ldots<j_N} a^\dagger_{j_1} b^\dagger_{j_1} \ldots a^\dagger_{j_N} 
b^\dagger_{j_N} \ket{0} 
\label{eq:coboson_ground_state}
\eeq 
It is straightforward to notice that this ansatz for the ground state cannot 
capture at all the effects of the effective repulsion appearing in the 
Hamiltonian for the ground manifold of the $N$ particles. Indeed, it is 
entirely determined by the hopping term, since this is the only term in the 
single-pair Hamiltonian. Nevertheless, one could expect that this is 
still a good approximation in the limit of low densities, for which coboson 
theory was developed. In that limit, two cobosons are anyway very unlikely to 
be found next to each other, so corrections due to repulsion may be negligible. 
However, this turns out not to be the case. 

\begin{figure}
 \includegraphics[width=\columnwidth]{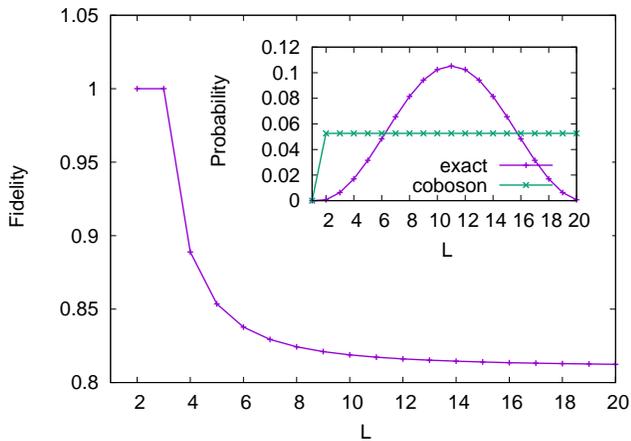}
 \caption{\label{fig:1D} Fidelity between the analytical ground state and the coboson ansatz for 
two pairs in the very strongly bound limit, in a 
one-dimensional lattice with $L$ sites and periodic boundary conditions. Inset: probability to find 
a pair in each site of the lattice conditional on having found one pair in the first site, for 
$L=20$, evaluated over the analytical ground state (\ref{eq:true ground state}) and over the 
coboson ansatz, displayed with violet~$+$ and turquoise $\times$ symbols respectively.}
\end{figure}

For the particular case $N=2$, 
which is the first non-trivial scenario to which coboson theory can be applied 
and for which the analytical solution of the problem is given in Eq. 
(\ref{eq:true ground state}), one can 
calculate the fidelity between the exact ground state and the coboson ansatz. 
This is defined as:
\beq
\mathcal{F}(\ket{N}, \ket{G}_N) = |\braket{N}{G}_N|^2
\eeq
and one finds that $\mathcal{F}$ actually decreases with the number of sites. 
The general 
calculation is cumbersome, but the limit $L\to\infty$ is particularly simple 
because a continuum limit can be taken turning sums over sites into integrals. 
For an infinite number of sites, i.e. for pair density tending to zero, the 
fidelity approaches the value 
$\mathcal{F}_\infty = 8/\pi^2\simeq 0.81$. The behaviour of the fidelity as a function of $L$ is 
shown in Fig. \ref{fig:1D}.

Numerical calculations show that the fidelity between the 
coboson ansatz and the ground state obtained numerically also 
decreases with the number of sites for larger values of particles. This is 
reasonable since the coboson ansatz is not expected to improve as the number of 
particles gets larger \cite{PC2018}; indeed, in coboson theory the dominant terms in an 
expansion in powers of the particle density are determined by the 
solutions of the problems of one and two pairs \cite{combescot_2008b, combescot_2011}. This is why 
we restrict our report to the most significant case of two pairs. 

We remark that it is possible to compute analytically the energy 
associated with the coboson ansatz for the ground state, and its value does 
converge to the right ground energy as the number of sites increases. Indeed, the ground-state 
energy of the effective Hamiltonian for two pairs is equal to:
\beq
E_{G, N=2} = -2 U_0 - 4\frac{J^2}{U_0} 
\cos^2\left(\frac{\pi}{2(L-1)}\right)
\eeq
whereas the coboson ansatz leads to the result:
\beq
E_{G, N=2} \simeq \langle 2| H_g^{(2)}|2 \rangle = -2 U_0 - \frac{4 J^2}{U_0} + \frac{4 J^2}{U_0} 
\frac{1}{L-1} \,.
\label{eq:energy 1}
\eeq
Thus, the two expressions approach each other as $L$ tends to infinity. However, 
this is merely due to the fact that the contribution of the interactions to the 
energy goes to zero as the pair density becomes negligible. Indeed, for large $L$ the ground-state 
energy of two pairs tends to twice the value of the single-pair ground-state 
energy, Eq. \eqref{eq:energy one pair}, as one would expect. 

The reason for the bad performance of the coboson ansatz can be traced back to the long-range 
character of 
the correlations between pairs in the true ground state (\ref{eq:true ground 
state}). Given the position of one of the pairs, the probability to find the 
other at a distance $d$ is proportional to $\sin^2[\pi 
(d-1/2)/(L-1)]$, i.e. it varies smoothly from zero for short distances to the 
maximum value when the pairs are at opposite positions in the chain. The 
coboson ansatz, on the contrary, predicts a flat probability distribution with 
equal probabilities for all non-zero distances between pairs. The contrast between the spatial 
correlations present in these two states is illustrated in the inset of Fig. \ref{fig:1D}, which 
shows the probability to find one pair as a function of position conditioned on the fact that the 
other pair is located in the first site.

\section{The two-dimensional case}\label{sec:2D}

The failure of coboson theory to give a good approximate description of the 
ground state in the 1D toy model studied is non-trivial, since the 
interactions are very short ranged, pairs are strongly bound, and 
the single-pair ground state can contain arbitrarily high entanglement. The 
reason why coboson theory is not applicable in this model seems to 
be that even for very low densities the ground state of two bounded pairs presents 
infinite-range correlations between the pairs. But this, in general, cannot be 
known until one solves for the ground state, which is exactly what one 
wishes to avoid by using the coboson ansatz. This naturally leads to the 
question, is there a key feature of the model that allows one to identify when 
coboson theory starts failing? Some rapid conjectures come to mind: the failure 
can be due to the 1D character of the model, the impenetrability of pairs (which can never cross 
each other), or the discretization of 
space. In this Section we analyze some of these possibilities.

\begin{figure}
 \includegraphics[width=\columnwidth]{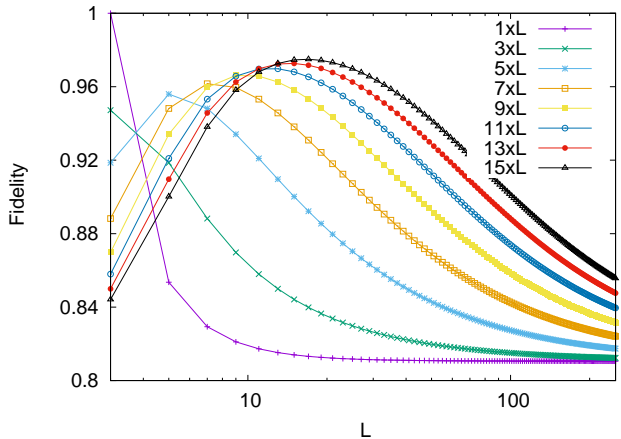}
 \caption{\label{fig:nxS} Fidelity between the ground state found 
numerically and the coboson ansatz for two pairs in an $n\times L$ lattice with torus 
boundary conditions, and where the effective tunneling 
strength is the same in both directions. For fixed $n$, in all cases the 
fidelity decreases with $L$.}
\end{figure}

We thus consider the simplest extension of the previous model: another 
lattice with $n\times L$ sites, so pairs can go around each other. For 
definiteness, we take periodic boundary conditions 
in both directions (i.e. a torus geometry). The basis of states of one fermion of 
either kind, or of one composite boson in the strongly bound regime, is given by 
the set of possible positions $\ket{(j_x,j_y)}$ with $j_x =1, \ldots, L$ and $j_y 
=1, \ldots, n$. The Hamiltonian is analogous to the one in (\ref{eq: eff H}), 
except that interactions and tunneling can involve any pair of neighbouring 
sites. Without breaking the translational invariance that is key for the simple 
form of the coboson ansatz and for the maximum entanglement between pair components, one 
can take two different effective tunneling 
constants, one for each direction, i.e. $J_\nu^{\rm eff} = J_\nu^2/U_0$ 
with $\nu = x, y$. 

In order to obtain the ground state of the model numerically, we exploit the translational
invariance restricting to the zero-momentum subspace. By doing this we are able to treat larger 
systems. The results for the 
fidelity between the numerical ground state and the coboson 
ansatz are shown in Fig.~\ref{fig:nxS}, where for definiteness we take 
$J_x=J_y$. In each of the curves, we set $n$ fixed and decrease the 
pair density by increasing the value of $L$. Once more, the coboson ansatz 
fails to reproduce the features of the ground state in the limit of low 
densities. One can observe from the figure that, for each value of $n$, at 
first the fidelity increases with $L$, reaching a maximum value when $L\simeq 
n$. From that point, the fidelity decreases with $L$, and the system behaves as 
one-dimensional. We have also analyzed cases where 
the tunneling constant is larger in one of the directions than in the other, 
and found that the values of the fidelity may vary with this choice but the 
decreasing trend of the fidelity for large $L$ is general. 

\begin{figure}
\includegraphics[width=\columnwidth]{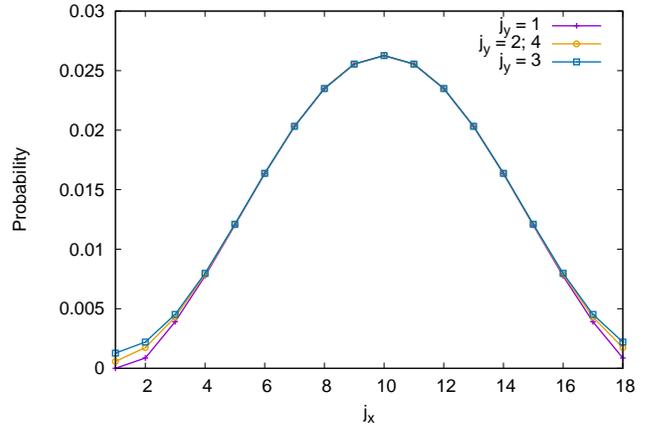}
 \caption{\label{fig:correlations_1D} Correlations between two pairs in the 
ground state of 
the $4\times 18$ lattice with equal tunneling in both directions. Given the 
position of one pair at site (1,1), the plot shows the 
probability to find the other pair as 
a function of the position in the same sublattice, $\mathrm{j_y}=1$ (violet dashes) 
and in the other three sublattices ($\mathrm{j_y}=2,3,4$, where the cases 2 and 4 are 
equal for symmetry reasons). 
Apart from the region which is closest to (1,1), the probabilities for all four 
sublattices are very similar and resemble the 
sinusoidal distribution found for the $1\times L$ lattice.}
\end{figure}

Once more, the behaviour of the fidelity can be understood in terms of the 
presence of long-range correlations in the positions of the pairs. Indeed, 
also for the $n\times L$ lattice one observes a pattern in the relative 
positions that resembles the $1\times L$ case. As an illustration, Fig. 
\ref{fig:correlations_1D} displays, for the case of equal 
tunnelings in both directions, the probability to find one pair relative to the 
position of the other for a $4\times 18$ lattice, showing that the solution of 
this model also exhibits strong and long-ranged spatial correlations between the 
two pairs.

\begin{figure}
 \includegraphics[width=\columnwidth]{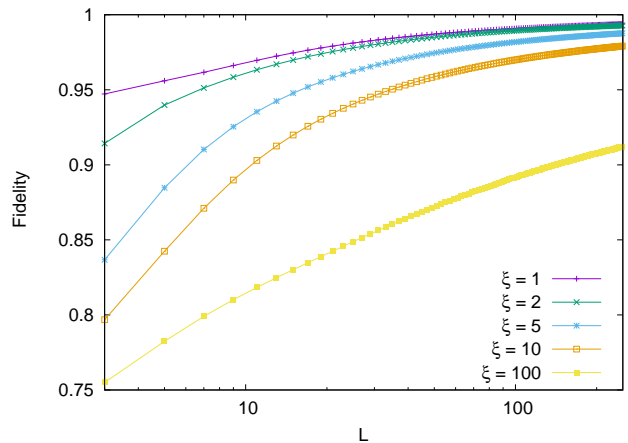}
 \caption{\label{fig:SxS} Fidelity between the ground state found numerically 
and the coboson ansatz for an $L\times L$ torus, as a function of $L$. The 
anisotropies are, from top to bottom, given by: $J_x^{\rm eff}/J_y^{\rm 
eff}=\xi$, for $\xi=$ 1, 2, 5, 10, 100.} 
\end{figure}

\begin{figure}
 \includegraphics[width=\columnwidth]{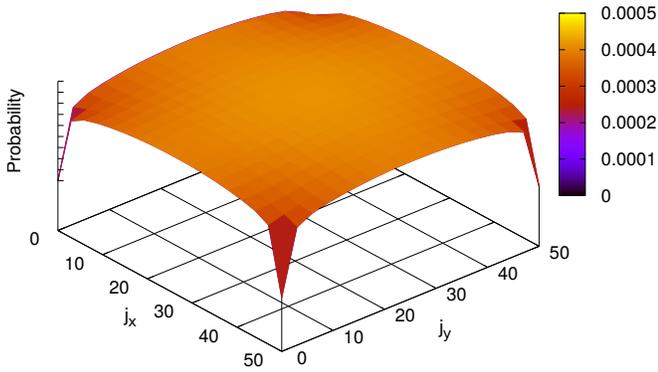}
 \caption{\label{fig:SxS_prob} Probability of finding a pair as a function of 
position given that there is a pair on site $(1,1)$, for an isotropic lattice with 51x51 
sites. 
As one moves away from the occupied site the probability becomes relatively 
flat, allowing for a good description in terms of the coboson ansatz.} 
\end{figure}

From Fig. \ref{fig:nxS} one can also observe that the maximum of the 
fidelity for each fixed $n$, found for $L\simeq n$, increases 
as a function of $n$. This suggests that the coboson ansatz is satisfactory 
in the truly two-dimensional case, i.e. when the low-density limit 
corresponds to a lattice size that increases in both dimensions. Fig. 
\ref{fig:SxS} 
shows the results for the fidelity in an $L\times L$ lattice as a function of 
$L$, for several cases corresponding to equal or different tunneling strengths 
in each direction. As 
can be seen from the figure, in this case the description of 
the ground state in terms of the coboson ansatz improves as the number of 
sites is increased approaching the very dilute limit. 

We note that all the cases plotted in Fig. \ref{fig:SxS} display a 
fidelity which increases with decreasing density, and the curves seem to 
asymptotically approach unit fidelity. However, the convergence is very slow 
and it strongly 
depends on the degree of anisotropy. This has a simple interpretation: for each 
finite value of $L$, if the tunneling in one of the directions is sufficiently 
large, there is a strong effective repulsion between pairs along that direction and each 
array of sites behaves effectively as a single cell. This makes the 
correlations in pair positions equivalent to the one-dimensional case. Nevertheless, 
fixing the values of the tunneling strengths and increasing sufficiently the 
value of $L$, the two-dimensional behaviour is always recovered, with a 
characteristic correlation length along each direction that depends on the 
corresponding tunneling strength.

For comparison with the previous cases, Fig. \ref{fig:SxS_prob} shows the 
correlations in the positions of the two pairs for a  two-dimensional lattice with $51\times 51$ sites; more 
precisely, the probability distribution for the position of the second pair is 
displayed 
conditioned on the first pair being found in site (1,1) and for the case when 
$J_x=J_y$. It can be seen that apart from a 
small region around the first site, the probability becomes relatively flat, 
which explains the good agreement with the coboson ansatz. These results indicate that the failure 
of the coboson ansatz observed in the previous Section can be associated with the one-dimensional 
character of the model leading to long-range spatial correlations.

\section{Conclusions}\label{sec:conclusion}

We have studied a toy model consisting of composite bosons strongly bound and 
tunneling along sites in a discrete lattice. Since we restrict to the case 
when the particles always tunnel in pairs, the model is equivalent to 
a hard-core Bose-Hubbard model, which is in turn equivalent to the Heisenberg model. The 
analysis of the one-dimensional case for two pairs has shown that the fidelity 
between the true ground state and the coboson ansatz decreases as the number of 
sites is increased. In the predictions of coboson theory for the case of $N$ 
pairs the dominant terms in an expansion with respect to pair density are given 
by the cases $N=1$ and $N=2$ \cite{combescot_2008b, combescot_2011}. This means 
that whenever the coboson ansatz fails 
to provide an appropriate description of the system for two pairs, it will also 
fail for higher numbers. Since the translational symmetry of the model studied 
makes the ground-state entanglement between pair constituents maximum, we 
conjecture that the coboson ansatz cannot generally be expected to faithfully 
describe one-dimensional discrete models. We related this with the presence of 
long-range correlations, and verified that this failure is also found in a 
slightly more complex model with an $n\times L$ lattice, where the low-density 
limit is taken for fixed $n$ and increasing $L$. The fact that the fidelity 
between the true ground state and the coboson ansatz decreases with decreasing 
density was also observed for models where the tunneling constant was different 
in the two directions.

The same analysis was carried out for a two-dimensional model 
corresponding to an $L\times L$ lattice with full translational symmetry. The 
results for a system of two pairs in this case show that the fidelity between 
the numerically found ground state and the coboson ansatz improves as the 
density is decreased, and it seems to approach the ideal unit value as the 
number of sites approaches 
infinity. We note, however, that the fidelity is strongly 
dependent on the degree of anisotropy of the model. Indeed, in systems 
where the two tunneling constants are very different the coboson ansatz is 
markedly less satisfactory than in isotropic models with the same number of 
sites.

Our study reveals a new aspect which is relevant for the 
fundamental understanding of when pairs of fermions are expected to behave 
approximately as elementary bosons, but which has received little attention so far. 
On top of a high amount of entanglement between 
constituents of a single pair, and a short-range character of the interactions 
so that for low densities pairs can be regarded as effectively independent, we observe a 
strong impact of the dimensionality of the system. One-dimensional models, 
namely lattices where the number of sites in one dimension is much larger than 
in the other, tend to develop long-range correlations in the positions of the 
pairs which cannot be captured by the coboson ansatz. On the contrary, 
truly two-dimensional lattices display a behaviour where the pairs can 
be approximately described as cobosons, with a fidelity that increases as the 
system becomes more dilute.

We note once more that our analysis restricts to the very strongly bound limit, such that 
the components of a pair always tunnel together and are always found in the same site. Systems of more loosely bound pairs are certainly of interest and can illustrate the gradual appearance of effective bosonic behaviour. For instance, the one-dimensional extended Hubbard model, with a tunable nearest-neighbour interaction, has been the focus of  \cite{lasmar_2019}. Here, however, we consider only the limiting case of very bound pairs because it is the 
one where the conditions which are normally expected to render the coboson ansatz valid are best satisfied.

It is important also to stress for clarity that our results are not directly 
connected with the well-known lack of condensation of a gas of non-interacting 
bosons at finite temperature and in the thermodynamic limit for less than 
three dimensions. Our models are studied at zero temperature and for finite 
system sizes. This means there is always a finite gap between the ground and the
excited states for a single coboson. In an actual experiment, however, the 
temperature can never be truly zero, so that thermal effects might mask those 
purely due to Pauli exclusion. The analysis of thermal states of composite 
bosons is a delicate task due to the overcompleteness of the coboson basis 
\cite{combescot_2011}, and lies beyond the scope of the present work.

Our results thus show that the coboson ansatz must be used with caution in situations where its 
validity has not been tested. We hope that our analysis will spark further 
interest in the understanding of the very relevant question of when composite 
particles can be treated as elementary.

\section*{Acknowledgments}

We are grateful to Elo\'isa Cuestas for discussions on the model and the 
coboson ansatz. This work used Mendieta Cluster from CCAD-UNC, which is part of SNCAD-MinCyT, Argentina.
 P.C. acknowledges Consejo Interuniversitario Nacional for the Beca Est\'imulo a las Vocaciones 
Cient\'ificas, call 2017. P.A.B. gratefully acknowledge support by the Conselho Nacional de Desenvolvimento Cient\'ifico e Tecnol\'ogico do Brasil and by the Spanish MINECO project FIS2014-59311-P (co-financed by FEDER). C.C. acknowledges funding from grant PICT-BID 2015-2236.


\end{document}